\documentclass[mathleft
]{an}
\usepackage{graphicx}
\usepackage{times}
\usepackage{amsmath}
\begin{document}

\Pagespan{789}{}
\Yearpublication{2010}%
\Yearsubmission{2010}%
\Month{11}%
\Volume{999}%
\Issue{88}%

\title{Convective Core Size Determination Using Asteroseismic Data}

\author{V. Silva Aguirre\inst{1}\fnmsep\thanks{Corresponding author:
  \email{vsilva@mpa-garching.mpg.de}\newline}
\and  J. Ballot\inst{2}
\and  A. Serenelli\inst{1}
\and  A. Weiss\inst{1}
}
\titlerunning{Convective core size}
\authorrunning{V. Silva Aguirre et al.}
\institute{
Max Planck Institute for Astrophysics, Garching, Germany
\and 
LATT/Observatoire Midi-Pyr\'en\'ees, Toulouse, France}

\received{}
\accepted{}
\publonline{}


\abstract{%
As it is long known, the presence of a convective region creates a discontinuity in the chemical profile of a star, which in turn translates into a sharp variation of the adiabatic sound speed. This variation produces an oscillatory behavior of the pulsation frequencies, to which low degree p-modes are sensitive. We investigate the possibility of detecting the signature related to the presence of a convective core in the frequency spectrum of low-mass stars by means of suitable frequency combinations (such as separations and ratios).}

\maketitle

\section{Introduction}\label{intro}
Asterosesimology has shown to be the most powerfull tool to study the inner properties of stars through the observation of stellar oscillations. Many efforts have been conducted to understand the interior physics of stars by understanding the influence that different physical phenomena have on the frequency spectra. In particular, it is known that sharp variations on the adiabatic sound speed produce an oscillatory signal in the p-mode spectra whose period is related to the position of such a discontinuity (e.g. \cite{sv88}, \cite{ap94}). Bearing this in mind, in this study we apply different asteroseismic tools to a computational model of a star containing a convective core and try to extract its position by means of the expected signatures on the frequency spectra.

\section{Signature of the convective core}\label{theo}
During its main sequence evolution stars with masses higher than $\sim$1.1 M$_\odot$ develop a convective core, whose extension is determined by the temperature stratification on the interior. Since the timescale for mixing of elements in this convective region is much shorter than the nuclear timescale, the core is believed to be homogeneously mixed in comparison with the radiative envelope, and a discontinuity in density appears in the edge of the fully mixed core. This sharp variation in density translates into a discontinuity in the adiabatic sound speed, to which p-modes are sensitive. Theoretically, the imprint left in the frequency spectrum of the acoustic modes is the appearance of an oscillatory signal, which is related to the location of the region of sharp variation within the star. In particular, the period of this oscillation relates to the exact radial coordinate where the discontinuity is located, and therefore to the travel time of the propagating wave throughout the acoustic cavity (\cite{mm94}).\\ 
The travel time of a propagating wave is represented by the acoustic radius or the acoustic depth, which are alternative representations of each other and are given by:
\begin{align}\label{eqn:time}
t&=\int_{0}^{r}\frac{\mathrm{d}r}{c_s}&
\tau&=\int_{r}^{R}\frac{\mathrm{d}r}{c_s},
\end{align}
where $c_s$ is the adiabatic sound speed. If the location of the density discontinuity is given by, say, $r_1$ in radial coordinates, and that same position is represented in acoustic radius and acoustic depth by $t_1$ and $\tau_1$ respectively, the period of the oscillation induced by this sharp variation is given by 1/(2$t_1$) and 1/(2$\tau_1$). 

\section{Frequencies, separations, and ratios}\label{seismic}
In order to study the inner structure of a star using asteroseismology, we need to find the appropriate tool which allows us to probe the desired region of the star an extract the required information. Different combinations of low-degree p-modes have been suggested as suitable probes of the physical characteristics of a star (e.g. \cite{jcd84}). The most commonly used are the so called large and small frequency separations, defined as:
\begin{align}\label{eqn:diff}
\Delta_{l}(n)&=\nu_{n,l}-\nu_{n-1,l}\\
d_{l,l+2}(n)&=\nu_{n,l}-\nu_{n-1,l+2},
\end{align}
where $\nu_{n,l}$ is the mode frequency of angular degree $l$ and radial order $n$. However, these combinations are affected by the outer layers of the star where turbulence in the near-surface is still poorly understood, and also changes in the adiabatic sound speed occur due to the presence of the convective envelope and the helium second ionization zone (e.g. \cite{jb04}). For the purpose of our study, we need to isolate from surface contamination the signal arising from the interior of the star in order to properly quantify the effects in the frequency spectra of a discontinuity in the chemical profile outside of the convective core. Therefore, we consider here the smoother 5 points small frequency separations and the ratio of small to large separations, quantities which are mainly determined by the inner structure of the star (\cite{rv03}) and are constructed as:
\begin{equation}\label{eqn:d01}
d_{01}=\frac{1}{8}(\nu_{n-1,0}-4\nu_{n-1,1}+6\nu_{n,0}-4\nu_{n,1}+\nu_{n+1,0})
\end{equation}
\begin{equation}\label{eqn:d10}
d_{10}=-\frac{1}{8}(\nu_{n-1,1}-4\nu_{n,0}+6\nu_{n,1}-4\nu_{n+1,0}+\nu_{n+1,1})
\end{equation}
\begin{align}\label{eqn:rat}
r_{01}&=\frac{d_{01}(n)}{\Delta_{1}(n)}&
r_{10}&=\frac{d_{10}(n)}{\Delta_{0}(n+1)}.
\end{align}
The small frequency separations have already been used to identify the location of the convective envelope and the helium second ionization zone (HeII) in the sun by applying them to observational data (\cite{ir09}), while it has been shown that the ratios fairly cancel out the influence in the frequencies of the outer layers (\cite{ir05}). We will explore here their potential in a more massive model than the sun to extract the signature produced by the sharp variation in the adiabatic sound speed outside of the convective core.

\section{Model computation}\label{model}
\begin{figure}[t]
{\includegraphics[width=\columnwidth]{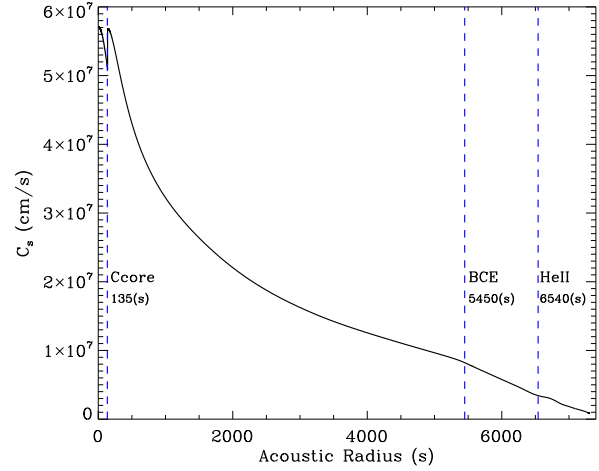}}
\caption{Adiabatic sound speed profile for the 1.5 M$_\odot$ model (solid black line). Also plotted in blue dashed lines are the location of the convective core boundary (Ccore), the base of the convective envelope (BCE) and the helium second ionization zone (HeII).}
\label{Ad_sound}
\end{figure}

\begin{figure}[t]
{\includegraphics[width=\columnwidth]{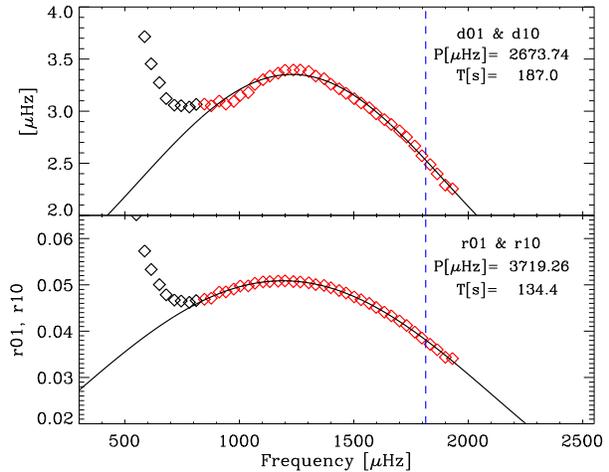}}
\caption{Small separations (upper panel) and ratios of small to large separations (lower panel). The frequencies marked in red have been used for the sine wave fitting (black solid line), and the period of the sine wave with its corresponding acoustic radius are also shown. The vertical blue dashed line corresponds to the acoustic cut-off frequency. The value of the period and its corresponding acoustic radius are on the upper-right side of each panel. See text for details}
\label{ratios}
\end{figure}

We have used a 1.5 M$_\odot$ model at solar metallicity constructed with GARSTEC (Garching Stellar Evolution Code, \cite{ws08}) during its main sequence evolution. Its central hydrogen content is Xc$\sim$ 0.25 and nuclear burning has produced a discontinuity in the chemical composition outside the fully mixed convective core. The adiabatic sound speed profile of the model is shown in Fig. \ref{Ad_sound}, where the location of the convective core, the convective envelope and the HeII ionization zone are also depicted.
We calculated the frequencies for the model using ADIPLS (Aarhus Adiabatic Oscillation Package, \cite{jcd08}) and constructed the small separations and ratios defined in Eqs. \ref{eqn:d01}, \ref{eqn:d10} and \ref{eqn:rat}. Both quantities are plotted in Fig. \ref{ratios}, where a clear long period oscillatory component is observed. As explained in Sect. \ref{theo}, the period of this oscillatory imprint in the frequency spectra can provide us with information about the location of the sharp discontinuity producing it, so we have done a non-linear least squares sinusoidal fit to the ratios and differences, and calculated the period of the obtained sine wave. The result of these fits are also plotted in Fig. \ref{ratios}, where the value obtained for the period of each curve is given on the upper-right side of each panel. We find that the long period oscillation observed is produced at a position of approximately $\sim$187 s for the small frequency separations and at $\sim$134 s for the frequency ratios, both values being in the vicinity of the convective core location ($\sim$135 s, see Fig. \ref{Ad_sound}). The results are summarized in Table \ref{results}, where the estimated position of the convective core extracted using the seismic tools is shown in mass and radius coordinates. We see that the ratios provide a more accurate location than the small separations, probably because the small separations still retain information from the outer parts of the star which is not being completely cancelled by this frequency combination (as it was already shown for the sun, see \cite{ir09}).
\begin{table}
\centering
\caption{ Convective core location -- in terms of acoustic radius (in seconds), fraction radius and fraction mass -- from the model, and the estimated location from the frequency combinations.}
\label{results}
\begin{tabular}{cccc}\hline
 & $t$ & m(r)/R & r/R\\ 
\hline
Model & 135.0 & 0.065 & 0.058\\
dd$_{01}$ \& dd$_{10}$ & 187.0 & 0.132 & 0.081\\
r$_{01}$ \& r$_{10}$ & 134.4 & 0.065 & 0.058\\
\hline
\end{tabular}
\end{table}

\section{Turning points}\label{ret_points}
\begin{figure}[t]
{\includegraphics[width=\columnwidth]{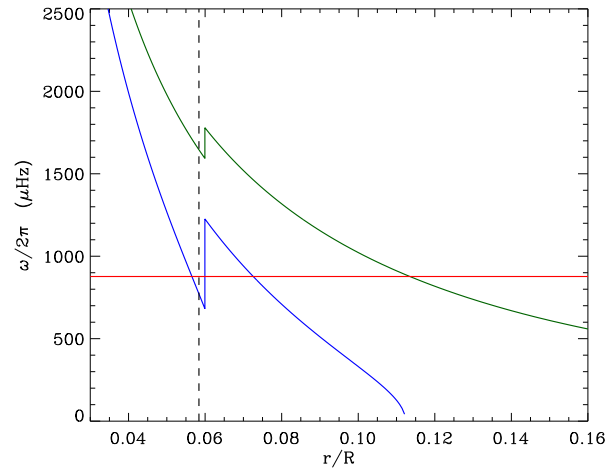}}
\caption{Characteristic acoustic frequency for the $l=1$ modes. Blue line shows the calculation made including the gravitational potential, while the green line depicts the one ignoring it. Black dashed line represents the position of the Convective Core, while the solid red line shows the frequency value of the lowest $l=1$ mode used to perform the sinusoidal fit described in Sect. \ref{model}. See text for details.}
\label{points}
\end{figure}

The information about the interior of the star that each acoustic mode carries depends in the region of the start it probes, and therefore in the acoustic cavity where it propagates. The radial ($l=0$) modes propagate down to the center of the star, but the situation for the $l=1$ modes is not so simple. In a first approximation, and neglecting the perturbation in the gravitational potential (\cite{tc41}), the propagation cavity of the p-modes can be estimated for each frequency value and the inner turning point calculated. This is the so called characteristic acoustic frequency, given by:
\begin{equation}\label{eqn:Sl}
\omega=\frac{c_s(r_t) L}{r_t},
\end{equation}
where $r_t$ is the radius where the turning point is located, and $L=(l+1/2)$ for $l=1$ modes. When the gravitational potential is taken into account, the fluctuations on it modify the dispersion relation of the incoming wave \cite{jj28} and the characteristic frequency is corrected as:
\begin{equation}\label{eqn:Sl2}
\omega^2=\frac{c_s(r_t)^2 L^2}{r_t^2}-4\pi G \rho(r_t),
\end{equation}
where $\rho(r_t)$ is the density in $r_t$. The inner turning point for each frequency value corresponds to the radius at which Eq. \ref{eqn:Sl} or Eq. \ref{eqn:Sl2} is satisfied. We calculated this values for the $l=1$ modes in both formulations to check if the modes used in the sinusoidal fit shown in Fig. \ref{ratios} actually penetrate the convective core. The obtained characteristic frequencies are plotted in Fig. \ref{points} as a function of radius. The green line represents the prescription from Eq. \ref{eqn:Sl}, while the blue solid line shows the result for Eq. \ref{eqn:Sl2}.
It can be seen in Fig.~\ref{ratios} that there is a clear change in the behavior of the small separations (and the ratios) around $\sim 800\:\mathrm{\mu Hz}$. In the frame of the Cowling approximation, this cannot be explained: only dipolar modes above $\sim 1700\:\mathrm{\mu Hz}$ should reach the core. However, by considering the fluctuations in the gravitational potential (Eq. \ref{eqn:Sl2}), $l=1$ modes above $\sim 800\:\mathrm{\mu Hz}$ can reach the core -- even if it is worth noting that below $\sim 1200\:\mathrm{\mu Hz}$, they must cross an acoustic wall (see Fig. \ref{points}). Thus this corrected Lamb frequency appears to be a suited approximation to know which modes can reach the core or not. For a future exploration, we will considered the results of \cite{mt06} who has developed a self-consistent theory for dipolar modes without the Cowling approximation.

\section{Conclusions and future development}
The diagnostic potential of the long known periodic component of frequency perturbations due to a discontinuity or steep gradient in the adiabatic sound speed has been used to estimate the position within the acoustic cavity of such sharp feature. In particular, we have estimated the location of the convective core using a sine wave fit to the large oscillation pattern observed in two different frequency combinations, showing that the accuracy in the position determination is higher when the ratios are considered than using the small separations. We intend to continue this study and extend it to other mass ranges and different types of discontinuities, such as the ones produced by overshooting and semiconvection (\cite{vs10}). We are aware that the potential of this diagnosis depends on the ranges of radial orders considered for the sine wave fitting of the data, for which a more elaborate analysis of the turning points of the dipole modes will be applied (e.g \cite{mt06}). Finally, we plan to complement this study by also applying other asteroseismic tools developed to detect small convective cores to our models (e.g \cite{cm07}, \cite{am06}), and explore the possibilities of detecting these mixed zones with the current data being obtained by the CoRoT and Kepler missions.

\end{document}